\documentclass[10pt]{article}

\usepackage{amsmath}
\usepackage{amssy
mb}

\usepackage{graphicx}

\usepackage{cite}

\usepackage{color} 

\usepackage[T1]{fontenc} 

\makeatletter
\renewcommand{\@biblabel}[1]{\quad#1.}
\makeatother

\date{}

\begin{document}

\begin{flushleft}
{\Large
\textbf{%
{Zipf's law for word frequencies: word forms versus lemmas in long texts.}
}
}
\\
\'Alvaro Corral$^{1,2,\ast}$, 
Gemma Boleda$^{3}$, 
Ramon Ferrer-i-Cancho$^{4}$
\\
\bf{1} 
Centre de Recerca Matem\`atica, Bellaterra, Barcelona, Spain
\\
\bf{2} 
Departament de Matem\`atiques, Universitat Aut\`onoma de Barcelona, Bellaterra, Barcelona, Spain
\\
\bf{3} 
Department of Translation and Language Sciences, Universitat Pompeu Fabra, Barcelona, Spain
\\
\bf{4} 
{Complexity and Quantitative Linguistics Lab}, Departament de Ci\`encies de la Computaci\'o, 
Universitat Polit\`ecnica de Catalunya, Barcelona, Spain
\\
$\ast$ E-mail: acorral@crm.cat
\end{flushleft}

\section*{Abstract}
{Zipf's law is a fundamental paradigm in the statistics of written and spoken natural language
as well as in other communication systems.
We raise the question of the elementary units for which Zipf's law
should hold in the most natural way, studying its validity for plain word forms
and for the corresponding lemma forms.
We analyze 
several long literary texts comprising four languages,
with different levels of morphological complexity.
In all cases Zipf's law is fulfilled, 
{in the sense that a power-law distribution of word or lemma frequencies is valid}
for several orders of magnitude.
{We investigate the extent to which the word-lemma transformation preserves two parameters of Zipf's law: the exponent and the {low-frequency} cut-off}. 
We are not able to demonstrate a strict invariance {of the tail}, as for a few texts 
both exponents deviate significantly, but we conclude that the exponents 
are very similar, despite the remarkable transformation that 
going from words to lemmas represents, considerably affecting all ranges of  
frequencies. {In contrast, the {low-frequency} cut-offs are less stable,
{tending to increase substantially} after the transformation.}
}

\section*{Introduction}

{Zipf's law for word frequencies is one of the {best known}
statistical regularities of language \cite{Zipf1972,Zanette_book}. 
In its {most popular} {formulation}, the law states that the frequency $n$ of the $r$-th most frequent word of a text follows 
\begin{equation}
n(r) \propto \frac 1 {r^{\alpha}},
\label{laprimera}
\end{equation}
where $\alpha$ is a constant
{and $\propto$ the symbol of proportionality}. 
However, Eq. (\ref{laprimera}) is not the only possible approach for modeling word frequencies in texts. 
One could also look at the number of {different} words with a given frequency {in} {a} text. 
In that case, 
{the probability $f(n)$ that a word has
frequency $n$ is {given by}}
\begin{equation}
f(n) \propto \frac 1  {n^{\gamma}},
\label{lasegunda}
\end{equation}
where $\gamma$ is a constant. 
The real values of $f(n)$ and $n(r)$ contain the information about the frequency of the words in a text, 
but $f(n)$ does it in a compressed fashion (given only the values of $f(n)$ such that $f(n)>0$, 
$n(r)$ is retrieved for any 
value of $r$).
In the {first} version of the law,
{$r$, the so-called rank of a word, acts as the random variable,}
and in the second version 
the random variable is the frequency of a word, $n$. 
In both cases, $\alpha$ and $\gamma$ are the exponents,
{related by \cite{Zanette_book}
{\begin{equation}
\gamma=1+\frac 1 \alpha.
\label{exponents_equation}
\end{equation}}}
{Usually, $\alpha$ is close to 1 and {then} $\gamma$ is close to 2.}

{{The relevance of Zipf's law for human language \cite{Miller1968a,Li1992b,Ferrer2009b}, 
as well as for other species' communication systems\cite{Suzuki2005a,McCowan2005a,Ferrer2012c}, 
has been the topic of a long debate.}
To some researchers, Zipf's law for frequencies is an inevitable consequence of the fact that words are made of letters (or phonemes): Zipf's law is obtained no matter if you are a human or another creature with a capacity to press keys sequentially \cite{Miller1968a,Li1992b} or to concatenate units to build words in a more abstract sense \cite{Suzuki2005a}. 
This opinion is challenged 
{by empirical values of $\alpha$}
that are not 
covered by simple versions of random typing \cite{Ferrer2005c}{, the dependence of $\alpha$ upon language complexity during language ontogeny \cite{Baixeries2012a},} 
and also {by} the large differences between the {statistics} defined on ranks (e.g., the mean rank) and the random typing experiments with parameters for which a good fit was claimed or expected \cite{Ferrer2009b}. }

{If the law is not inevitable, understanding the conditions under which it emerges or varies is crucial. Alterations in the shape and parameters of the law have been reported in child language \cite{Baixeries2012a,Piotrowski1995}, schizophrenic speech \cite{Piotrowski1995, Piotrowski2007a}, {aphasia \cite{VanEgmond2011a,Hernandez2013a},} 
{and large multiauthor texts \cite{Ferrer2000a,Petersen2012b,Gerlach2013a}}.
Despite intense research on Zipf's law in {quantitative} linguistics {and complex systems science}, 
little attention has been paid to the elementary units 
for which Zipf's law should hold. 
Zipf's law has been investigated in letters \cite{Naranan1993}
and also in blocks of symbols (e.g., words or letters) \cite{Egghe2000a}.
Here a very important issue that has not received enough attention 
{since the seminal work of Zipf}
is investigated in depth: {the} effect of considering word forms vs.\ lemmas in the presence, scope and parameters of the law
(a lemma is, roughly speaking, the stem form of a word; see below for a more precise definition).
Research on this problem is lacking as the overwhelming majority of empirical research 
has focused on word forms for simplicity 
(e.g., \cite{Baayen,Jayaram2008a,Tuzzi2009a,Petersen2012b,Gerlach2013a,Li2010a,Baixeries2012a}).}

{Thus, here we address a very relevant research question: 
does the distribution of word frequencies differ from that of lemmas? 
This opens {two} subquestions:
\begin{itemize}
\item
Does Zipf's law still hold in lemmas?
\item
Does the exponent of {the} law for word forms differ from that of lemmas? 
\end{itemize}
{
It is remarkable that Zipf himself addressed this problem
at a very preliminary level
(Fig. 3.5 in Ref. \cite{Zipf1972}), 
and it has not been until much {more} recently that
several researchers have revisited it.
{Baroni compared the distribution of ranks in a lemmatized version of the {\it British National Corpus} against the non-lemmatized counterpart and concluded, based upon a qualitative analysis, that both show essentially the same pattern \cite{Baroni2009}}.
Reference \cite{Kwapien}
{studied} one English text ({\it Ulysses}, by James Joyce) and
one Polish text;
for the former, the 
word and lemma rank-frequency relations 
were practically undistinguishable,
but for the Polish text
some differences were found:
the exponent $\alpha$ slightly increased
(from 0.99 to 1.03)
when going from words to lemmas
and a second power-law regime seemed to appear
for the highest ranks, 
{with exponent $\alpha$ {about} 1.5.}
Bentz et al. \cite{Bentz2014}, for a translation of the {\it Book of Genesis} into English,
pointed to a connection between
morphology and rank-frequency relations,
provided by an increase in the exponent $\alpha$
(from 1.22 to 1.29)
when the book was lemmatized and Mandelbrot's generalization of
Zipf's law was used in a maximum likelihood fit of $n(r)$.
{Finally, Hatzigeorgiu et al. \cite{Hatzigeourgiu2001a} analyzed the Hellenic National Corpus and found that 
the exponent $\alpha$ decreased when taking the 1000 most frequent units (from $\alpha = 0.978$ for the 1000 most frequent word forms to $\alpha = 0.870 $ for the 1000 most frequent lemmas). This decrease is hard to compare with the increases reported in Refs. \cite{Kwapien,Bentz2014} and the results presented in this article because it is restricted to the most frequent units.}

Our study will provide a larger scale analysis, 
with 10 rather long 
{single-author} texts 
(among them some of the longest novels in the history of literature)
in 4 different languages,
using state-of-the-art tools in {computational} linguistics
and {power-law fitting}. 
{%
The languages we study cover a fair range in the word-lemma ratio, 
from a morphologically poor language {such as} English to a highly inflectional language such us Finnish, 
with Spanish and French being in {between.}}}
In a previous study with a subset of these texts, some of us
investigated the dependence of word and lemma frequency distributions 
on text length \cite{FontClos_Corral}, but
no direct quantitative comparison was performed 
between the results for word and lemmas.
It will be shown here 
that the range of validity of Zipf's law 
[Eq. (\ref{lasegunda})] decreases 
when using lemmas; however, {we will show that}, while
the exponents obtained with word forms and lemmas do not 
{follow the same distribution, they maintain a very close and simple
relationship,}
suggesting some robust underlying mechanism. 
}

{We will study the robustness of Zipf's law concerning lemmatization
from the perspective of type frequencies instead of ranks. Ranks have the disadvantage of leading to a histogram or spectrum that is monotonically decreasing by definition. This can hide differences between real texts and random typing experiments \cite{Ferrer2009a}. 
The representation in terms of the distribution of frequencies $f(n)$ 
has been used successfully to show the robustness of Zipf's {exponents} as texts size increases: 
Although the shape of the distribution apparently changes as text length increases, a simple rescaling 
allows one to unravel a mold for $f(n)$ that is practically independent from text length \cite{FontClos_Corral}. 
In this article we investigate the extent to which $f(n)$ is invariant upon lemmatization. }
%
%
We restrict our analysis to single-author texts, more concretely literary texts.
This is because of the alterations in the shape and parameters of the distribution of word frequencies known to appear in large multi-author corpora \cite{Ferrer2000a,Petersen2012b,Gerlach2013a}.




\section*{Definitions}

Let us consider, in general,
a sequence composed of symbols that can be repeated.
We are studying texts composed by words, 
but the framework is equally valid for 
a DNA segment constituted by codons \cite{Mantegna_PRE_DNA}, a
musical piece consisting of notes \cite{Serra_scirep}, etc.
Each particular occurrence of a symbol is called 
a token, whereas the symbol itself is referred to as a type \cite{Baayen}.
The total number of tokens gives the sequence length, $L$
(the text length in our case), 
whereas the total number of types 
is the size of the {observed} vocabulary, $V$,
with $V \le L$.

In fact, although a sequence may be perfectly defined,
its division into symbols is, up to a certain point, arbitrary.
For instance, 
texts can be divided into letters, morphemes, 
etc., 
{but most studies in quantitative linguistics have considered the 
basic unit to be the} word.
{This is a linguistic notion that can be operationalized in many languages 
by delimiting 
{sets of letters} separated by spaces or punctuation marks.}
Nevertheless, the symbols that constitute {themselves} a sequence can be non-univocally related
to some other entities of interest, 
as it happens with the relationship between a word and its lemma.
A lemma is defined as
{a linguistic form that stands for or represents a whole inflectional morphological paradigm, 
such as the plural and singular forms of nouns or the different tensed forms of a verb. 
Lemmas 
are typically used as headwords in dictionaries.
For example, for a word type, {\it houses}, 
{the corresponding lemma type is
{\it house}.}}
Nevertheless, this correspondence is not always so clear \cite{Popescu_Altmann},
such that lemmatization is by no means a straightforward transformation.

{Using different texts, {we will check the validity of Zipf's law for lemmas, and} 
we will compare the statistics of {word forms} to the statistics of lemmas.
To gather statistics for lemmas, we will replace
each word in the text by its {associated} lemma,
and will consider the text as composed by lemmas.
To see the effect of this transformation, consider
for instance the word {\it houses} in {\it Ulysses}.
The number of tokens for the word {type} {\it houses} is 26, 
because {\it houses} occurs 26 times {in the book}.
However, the number of tokens for the corresponding lemma, {\it house}, is 198, 
because the lemma {\it house} (in all its nominal and verbal forms, 
{\it house, houses, housed}...) occurs 198 times.
The} relationship between the statistics of words and lemmas,
{and in particular the question of whether lemmas follow Zipf's law
 or not \cite{Zipf1972}},
{is not a trivial issue \cite{Popescu_Altmann}.}

{
In order to investigate the validity of Zipf's law in a text
{we count the frequency $n$ of all (word or lemma) types}
and}
fit the tail of the distribution of frequencies {(starting at some point $n=a$)} 
to a power law, {i.e.,}
$$
f(n) =\frac C {n^{\gamma}}, \, \mbox{ for } \, n\ge a,
$$
{with $\gamma >1$, $C$ the normalization constant,} and
disregarding values of $n$ below $a$. 
{The version of Zipf's law that we adopt has two parameters: 
the exponent $\gamma$ and the low-frequency cut-off $a$. 
We consider that Zipf's law is valid if a power law holds starting at $a$ 
and reaching at least two decades {up to the maximum frequency
(the frequency of the most common type)}. 
With these assumptions, we are adhering to the view of Zipf's law as an asymptotic property of a random variable \cite{Conrad2004a,Stumpf2012a}.}
 
{To fit this definition of the law we use a two-step {procedure} 
that first fits the value of $\gamma$ for a fixed $a$ and next evaluates the
goodness of the power-law fit from $a$ onwards;
{this is repeated for different $a$-values until the most satisfactory fit is found}. 
The resulting exponent is reported as $\gamma\pm\sigma$,
where $\sigma$ is the standard deviation of $\gamma$.
Our procedure is} similar in spirit to the one by Clauset et al. \cite{Clauset}, 
but it {can be shown to have} a better performance 
for continuous random variables \cite{Peters_Deluca,Corral_nuclear,Corral_Deluca}.
Indeed, Clauset et al.'s requirement for power-law acceptance 
seems to be very strict,
having been found to reject the power-law hypothesis
even for power-law simulated data \cite{Corral_nuclear}.
{Details of the procedure we use are explained in Ref. \cite{Corral_Deluca_arxiv};
this is basically the adaptation of the 
method of Ref. \cite{Corral_Deluca} to the discrete case.
The {\it Materials and Methods} Section provides a summary.}

\section*{Results}

We analyze a total of 10 novels {comprising four languages:}  English, Spanish, French, and Finnish,
see Table \ref{Tableone}.
In order to gather
enough statistics, 
we include
some of the longest novels ever written, {to our knowledge}.
{For the statistical analysis of lemmas}, 
we first perform an automatic process of lemmatization
using state of the art computational tools.
The steps 
comprise tokenization, morphological analysis, and morphological disambiguation, 
in such a way that, at the end, each word token is assigned a lemma.
See {\it Materials and Methods} for {further} details. 

\subsection*{{Zipf's law holds for both word forms and lemmas}}

Fig. \ref{Fig1a}(a) {compares} the results {before and after lemmatization} 
for the book {\it La Regenta}
{(in Spanish)}.
The frequency distributions $f(n)$ for words and for lemmas are certainly different,
with higher frequencies being less likely 
for words than for lemmas,
an effect that is almost totally compensated by {{\em hapax legomena} (types of frequency {equal to} one)},
where words have more weight than lemmas.
This is not unexpected, as the lemmatization process leads to 
less types (lower $V$), which {must} have higher frequencies, on average
(the mean frequency is $\langle n \rangle = L/V$).
{The reduction of vocabulary for lemmas (for a fixed text length)
has a similar effect to that of increasing text length;
in other words, we are more likely to see the effects of the exhaustion 
of vocabulary {(if this happens)} using lemmas rather than words.}
The {difference} in the counts of frequencies results in a tendency of $f(n)$ for lemmas
to bend downwards 
{as the frequency decreases towards the smallest values
(i.e., the largest ranks)}
in comparison with the $f(n)$ of words; 
this in agreement with Ref. \cite{Kwapien}.
%
{{Besides}, one has to take into account
that lemmatization errors are more likely for low frequencies, 
and then the frequency distribution in that domain
can be more strongly affected by such errors.
In any case, our main interest is for high frequencies,
for which the quantitative behavior 
shows a power-law tail for both words and lemmas}.
This extends for almost three {orders of magnitude},
{with exponents $\gamma$ very close to 2},
implying the {fulfillment} of Zipf's law 
(see Table \ref{Tabletwo}).

\begin{figure}[!ht]
\begin{center}
  \includegraphics[width=5in]{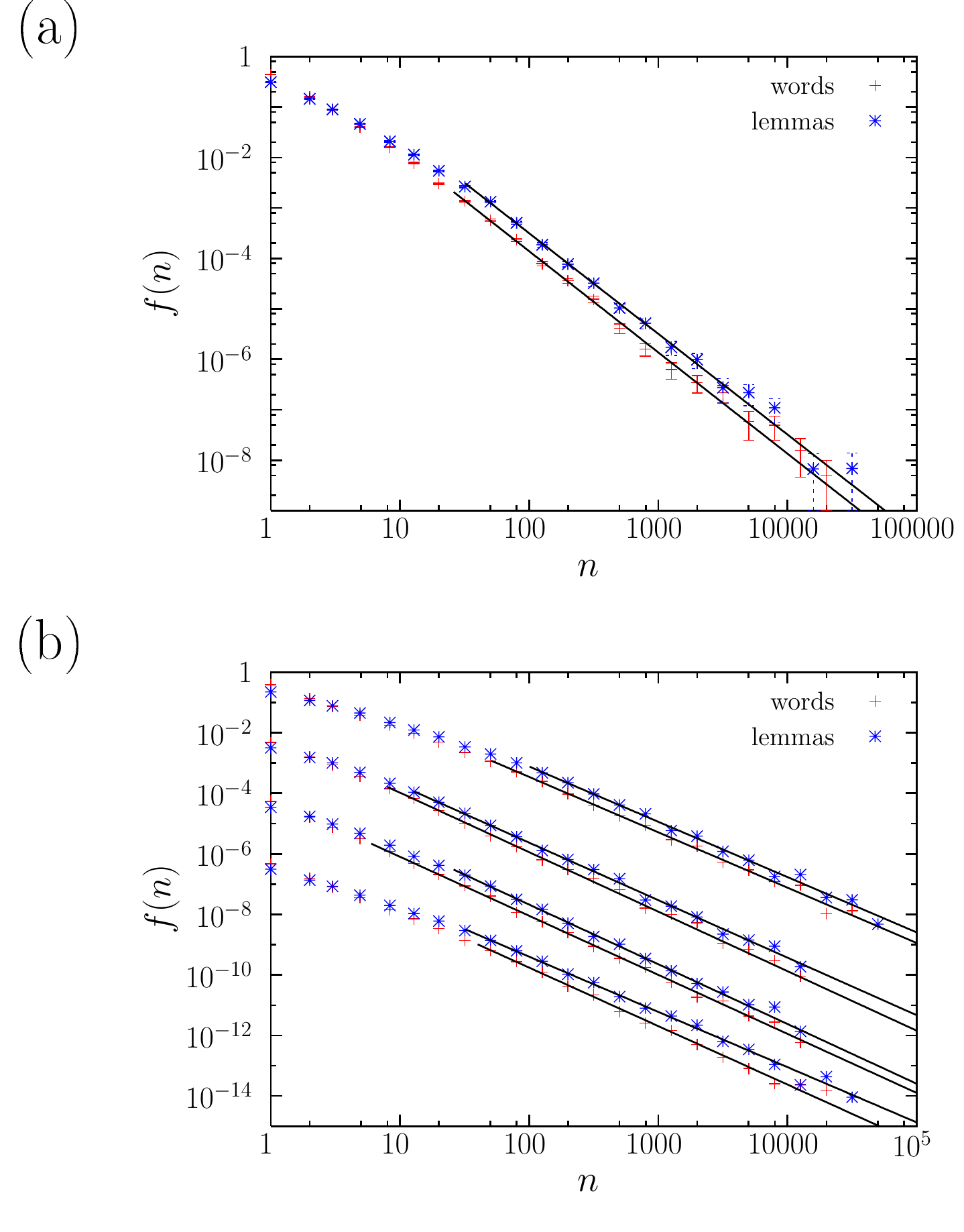}
\end{center}
\caption{
{(a) Probability mass functions $f(n)$ of the absolute frequencies $n$ of
words and lemmas in {\it La Regenta}, together with their fits.
(b) The same, from top to bottom, for {\it Clarissa, Moby-Dick, Ulysses}
(all three in English), and {\it Don Quijote} (in Spanish).
The distributions are multiplied by factors $1$, $10^{-2}$, $10^{-4}$
and $10^{-6}$ for a clearer visualization.
}  
}
\label{Fig1a}
\end{figure}

The rest of books analyzed show a similar qualitative behavior, 
as shown for 4 of them in Fig. \ref{Fig1a}(b).
{In all cases Zipf's law holds, both for words and for lemmas.}
The {power-law tail} exponents $\gamma$ range from 1.83 to {2.13}, see Table \ref{Tabletwo},
{covering {from} 2 and a half to 3 {and a half} 
orders of magnitude of the type frequency
(except for lemmas in {\it Seitsem\"{a}n veljest\"{a}}, with roughly only 2 orders 
of magnitude}).
{
For the second power-law regime reported in Ref. \cite{Kwapien}
for the high-rank domain of lemmas (i.e., low lemma frequencies), we only find it for 
the smallest frequencies (i.e., between $n=1$ and a maximum $n$)
in two Finnish novels, 
{\it Kev\"{a}t ja takatalvi} and {\it Vanhempieni romaani}
with exponents $\gamma=1.715$ and $1.77 \pm 0.008$, respectively.
These {values of $\gamma$} yield $\alpha=1.40$ and $1.30$ 
{(recall Eq. (\ref{exponents_equation}))}, 
{which one can compare} to the value obtained in Ref. \cite{Kwapien}
for a Polish novel (1.52). However, for the rest of distributions of lemma frequency, 
a discrete power law starting in $n=1$ is rejected no matter
the value of the maximum $n$ considered.
This is not incompatible with 
the results of Ref. \cite{FontClos_Corral},
{as a different fit and a different testing procedure was used there}.
Note that the Finnish novels yield the poorest statistics
(as their text lengths are the smallest),
so this second power-law regime seems to be significant
only for short enough texts.

\begin{figure}[!ht]
\begin{center}
\includegraphics[width=5in]{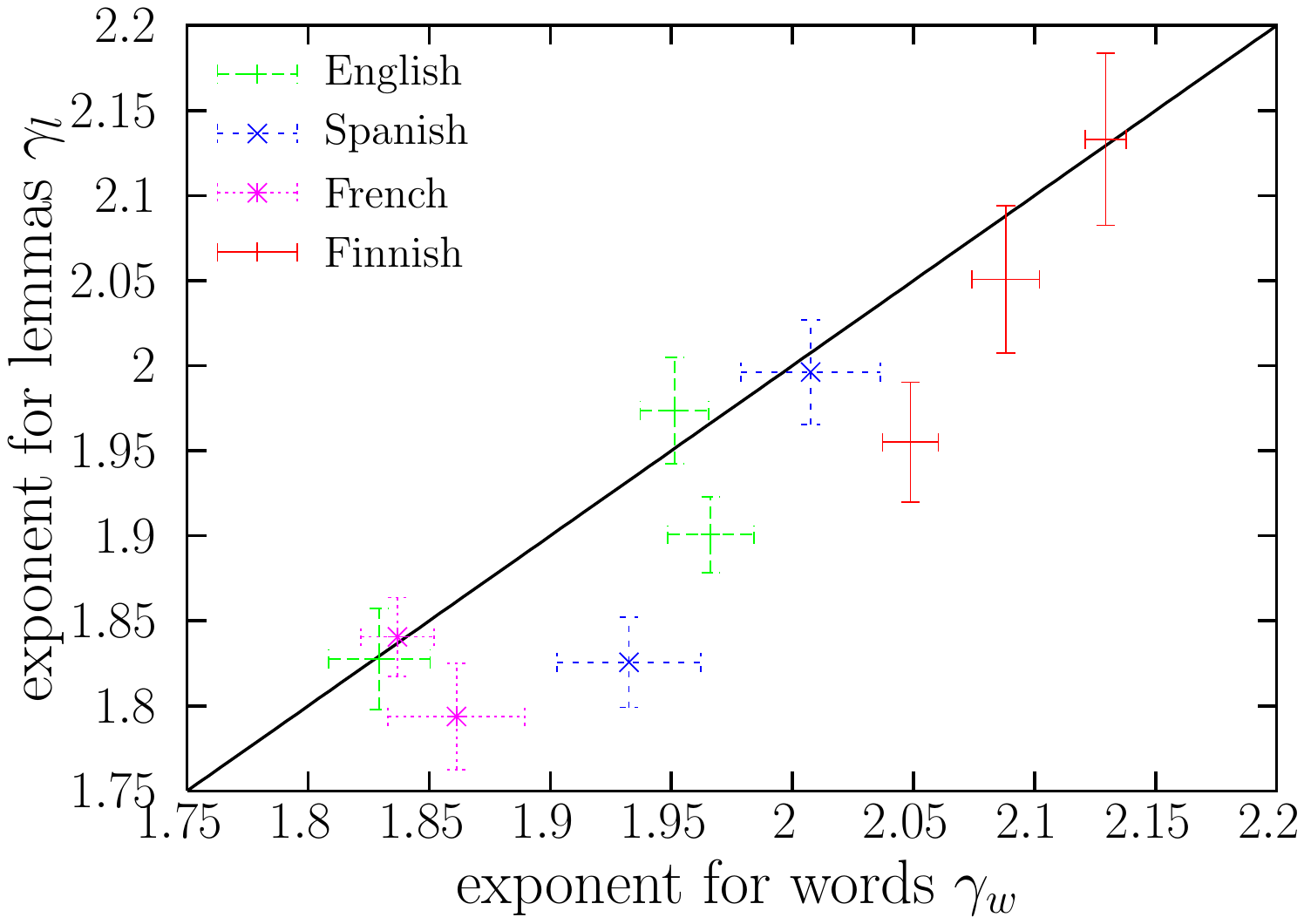} 
\end{center}
\caption{
$\gamma_l$ (the exponent of the frequency distribution of lemmas) versus $\gamma_w$ (the exponent of the frequency distribution of {word} forms). As a guide to the eye, the line $\gamma_l = \gamma_w$ is also shown (solid line).  
Error bars indicate one standard deviation. 
}
\label{Fig_extra}
\end{figure}

\subsection*{{The consistency of the exponents between word forms and lemmas}}

{In order to proceed with the comparison between}
{the exponents of
{the frequency distributions of} words ($w$) and lemmas ($l$), 
let us denote them as $\gamma_w$ and $\gamma_l$, respectively. 
{Those values are compared in Fig. \ref{Fig_extra}.}
{Coming back to} the example of {\it La Regenta}, it is remarkable that 
the two exponents do not show {a} noticeable difference 
(as it is apparent in Fig. \ref{Fig1a}(a)),
with values $\gamma_w=2.01\pm 0.03$ and $\gamma_l=2.00\pm 0.03$. 
}
Out of the remaining 9 texts, 4 of them give pairs of {word-lemma} exponents
with a difference of 0.02 or smaller.
This is within the error bars of the exponents, 
represented by the standard deviations $\sigma_w$ and $\sigma_l$ 
of the maximum likelihood estimations {of the exponents}; 
{{more precisely}, $|\gamma_w-\gamma_l| < \sigma_l$,
as can be seen in Table \ref {Tabletwo}.}
{
For the other 5 texts, 
the two exponents
are always in the range of overlap of 
two standard deviations, i.e., 
$|\gamma_w-\gamma_l| < 2(\sigma_w+\sigma_l)$.
}


{However, we should be cautious in drawing conclusions from these data.
If, {for a fixed book,} $\gamma_w$ and $\gamma_l$ were independent variables,
the standard deviation of their difference would be
$\sigma_d=\sqrt{\sigma_w^2 +\sigma_l^2}$, {according to {elementary probability theory} 
(\cite{Taylor1997a}, Chapter 3)};
however, independence cannot be ensured and 
we have $\sigma_d=\sqrt{\sigma_w^2 +\sigma_l^2-2 \mbox{cov}(\gamma_w,\gamma_l)}$,
where cov$(\gamma_w,\gamma_l)$ is the covariance of both variables,
{for a fixed book 
{(this covariance is
{different from the
covariance implicit in the Pearson correlation introduced below, 
which refers to all texts).}}}
Although the maximum likelihood method provides an estimation for
the standard deviations of the exponents (for a fixed text) \cite{Clauset,Corral_Deluca}, 
we {cannot} compute the covariance of the word and lemma exponents
{(for the total size of each text), 
and therefore we do not know the uncertainty in the difference between them.
{{This is is due to the fact that} we only have one sample {for each book} to calculate $\gamma_w$ and $\gamma_l$.}
If we could assume independence, we would obtain that already three books 
yield results outside the 95 \% confidence interval of the exponent difference
(given by $2\sigma_d$), see Table \ref{Tabletwo}.
This could be modified somewhat by the
Bonferroni and {\v{S}id\'ak} corrections for multiple testing 
\cite{Bland_Altman,Abdi_Bonferroni}.
Nevertheless, we expect a {non-zero} covariance between $\gamma_w$ and $\gamma_l$,
as the samples representing words and lemmas have some overlap
{(for instance, some word tokens remain the same after lemmatization)}, and
therefore the standard deviation
$\sigma_d$ should be smaller than in the independent case,
which leads to larger significant differences than
what the independence assumption yields.}
{{Conversely}, the standard deviations $\sigma_w$ and $\sigma_l$
of the maximum likelihood exponents are obtained assuming that $a_w$ and $a_l$
are fixed parameters, but they are not, and then the total uncertainties
of the exponents are expected to be larger than the reported standard deviations;
nevertheless, this is difficult to quantify.}
{Thus, the standard deviations we provide for the exponents have to be interpreted as  
some indication of their uncertainty but not as the full uncertainty,          
which could be larger.} 
{We conclude that
we cannot establish an absolute invariance of the value of the Zipf exponent
under lemmatization.}


Instead of comparing the word and lemma exponents book by book, 
{using the uncertainty for each exponent}, 
we can also deal with the whole ensemble of exponents,
{ignoring the individual uncertainties}.   
{We consider first a Student's $t-$test for paired samples
{to analyze the differences between pairs of exponents}.
This test, although valid for dependent normally distributed data
(and the estimations of the exponents are normally distributed), 
assumes that the standard deviations $\sigma_w$ and $\sigma_l$
are the same for all books, which is not the case, see Table \ref{Tabletwo}.
So, as a first approximation we apply the test and interpret its results with care.
The $t-$statistics gives $t=2.466$ {($p$-value=0.036)},
leading to the rejection of the hypothesis that there are no 
significant differences between the exponents. {These results 
{do not look like}
very surprising upon {visual} inspection of Fig. \ref{Fig_extra}: Most points $(\gamma_w, \gamma_l)$ lie below the diagonal, suggesting 
a tendency for $\gamma_l$ to have a lower value than $\gamma_w$.}
{But} we can go one step further with this test and consider the existence of one
outlier, removing from the data the book with the largest difference between
their exponents. In this case one needs to avoid introducing any bias in the
calculation of the $p$-value. For this purpose, we simulate the $t-$Student
distribution by summing rescaled normal variables in the usual way (see {\it Materials and Methods}),
and remove (in the same way as for empirical data) the largest value of the variables.
This yields $t=2.053$ and $p=0.075$, which {suggests} that the values of the
exponents are not significantly different, except for one outlier.
However, as we have mentioned, this test cannot be conclusive and other tests are necessary. 
}

{We realize that} $\gamma_w$ and $\gamma_l$ are clearly dependent variables
(when considering all books). 
{Their} Pearson correlation, a measure of linear correlation, is $\rho  = 0.913$ 
(the sample size is $\mathcal{N}=10$ and $p = 0.0003$ is the $p$-value of a two-sided test
{with null hypothesis $\rho =0$}).
{Note that this correlation is different to the one given above by $\mbox{cov}(\gamma_w,\gamma_l)$,
which referred to a fixed book.}
{Given this, we formulate three hypotheses 
about the relationship between the exponents. 
{The first hypothesis is that $\gamma_w$ and $\gamma_l$ are identically distributed 
for a given {text} (but not necessarily for 
{different texts,} different authors, or different languages).}
The second hypothesis is that 
$\gamma_w$ is centered around $\gamma_l$, 
i.e., the {conditional} expectation of $\gamma_w$ given $\gamma_l$ is $E[\gamma_w|\gamma_l] = \gamma_l$.
{This means that a reasonable prediction on the value of $\gamma_w$
can be attained from the knowledge of the value of $\gamma_l$.}
The third hypothesis is the symmetric of the second, namely that $\gamma_l$ 
is centered around $\gamma_w$, i.e., the {conditional} expectation of $\gamma_l$ given $\gamma_w$ 
is $E[\gamma_l|\gamma_w] = \gamma_w$. 
The second and third hypotheses 
are supported by the strong Pearson correlation between $\gamma_w$ and $\gamma_l$,
{but these two hypotheses are not equivalent \cite{Poirier1995a}}. }

{{We define $\bar  \gamma_w $ and $\bar  \gamma_l  $ 
as the {average} values of $\gamma_w$ and $\gamma_l$, 
respectively, in our sample of ten literary {texts}. 
The first hypothesis means that given a certain {text}, 
$\gamma_w$ and $\gamma_l$ are interchangeable. 
If $\gamma_w$ and $\gamma_l$ are identically distributed for a certain {text}, 
then {the} absolute value of the difference between the means
$|\bar  \gamma_w   - \bar  \gamma_l  |$ 
should not differ significantly from {analogous values} obtained by chance, 
i.e., flipping a fair coin to decide if $\gamma_w$ and $\gamma_l$ 
{remain the same} or {are} swapped within a book. 
As there are ten literary texts, there are $2^{10}$ possible configurations. 
Thus, one can compute numerically the 
$p$-value as the proportion of 
these configurations where $|\bar  \gamma_w   - \bar  \gamma_l  |$ 
equals or exceeds the original value. 
{This coin-flipping test is in the same spirit as Fisher's permutational test 
(\cite{Conover1999a}, pp. 407-416), 
with the difference that we perform the permutations
of the values of the exponents only inside every text. 
The application of this test}
reveals that $|\bar  \gamma_w   - \bar  \gamma_l  | = 0.035$,
{which is a
significantly large difference} 
(with a $p$-value $= 0.04$).
Therefore, we conclude that the first hypothesis does not stand,
{and therefore $\gamma_w$ and $\gamma_l$ are not identically distributed {within books}.
{This {seems} consistent with the fact that most points $(\gamma_w, \gamma_l)$ lay below the diagonal, {see} Fig. \ref{Fig_extra}. }
{However}, the elimination of one outlier (the text with the largest difference) 
leads to $p=0.08$,}  
{which makes the difference non-significant for the remaining texts.}
}
}


{The second hypothesis is equivalent to $E[\gamma_w/\gamma_l|\gamma_l] = 1$ 
and therefore this hypothesis is indeed that {the ratio} $\gamma_w/\gamma_l$ is mean independent of 
$\gamma_l$ {(the definition of mean independence in this case is
$E[\gamma_w/\gamma_l|\gamma_l] =$ constant $=E[\gamma_w/\gamma_l]$, 
(\cite{Poirier1995a}, pp. 67))}. 
Similarly, the third hypothesis is equivalent to 
$E[\gamma_l/\gamma_w|\gamma_w] = 1$ and therefore this hypothesis is indeed that 
$\gamma_l/\gamma_w$ is mean independent of $\gamma_w$. 
Mean independence can be rejected by means of a correlation test as 
mean independence needs uncorrelation 
(see Ref. \cite{Kolmogorov1956aa}, pp. 60 or Ref. \cite{Poirier1995a}, pp. 67).
A significant correlation between $\gamma_w/\gamma_l$ and $\gamma_l$ 
would reject the second hypothesis while a significant correlation between 
$\gamma_l/\gamma_w$ and $\gamma_w$ would reject the third hypothesis. 
Table \ref{mean_independence_table} indicates that neither 
the Pearson nor the Spearman {correlations are significant {(see {\it Materials and Methods}}), 
and therefore these correlation tests are not}
able to reject the second and the third hypotheses. 
Further support for the second and third hypotheses comes from linear regression. 
The second hypothesis states that 
$E[\gamma_w |\gamma_l] = c_1 \gamma_l + c_2$ with $c_1 = 1$ and $c_2 = 0$ 
while 
the third hypothesis states that 
$E[\gamma_l |\gamma_w] = c_3 \gamma_w + c_4$ with $c_3 = 1$ and $c_4 = 0$. 
Consistently, a standard linear regression {and subsequent statistical tests}
{indicate} that $c_1,c_3 \approx 1$ 
and $c_2,c_4 \approx 0$ {cannot be rejected} (Table \ref{linear_regression_table}). }

In any case, {to perform our analysis we have not taken into account} 
that the number of datapoints ($V$) and the {power-law}
fitting ranges are different for words
and lemmas, a fact that can increase the difference between
the values of the exponents
{(due to the fact that the detection of deviations from power-law behavior depends on the 
number of datapoints available)}. 
{In general,} the fitting ranges are larger {for} words than
for lemmas, due to the bending of the lemma distributions,
{see below}.
Another source of variation to take into account 
for the difference between the exponents
is, {as we have mentioned,} that the lemmatization process is not 
exact, which can lead to type assignment errors
and even to some words not being associated
to any lemma ({see the {\it Materials and Methods} Section for details).
}

Although,
{after the elimination of one outlier,
we are not able to detect differences between the exponents,}
{there seems to be a tendency for the lemma exponent
to be a bit smaller than the word exponent, as can be seen in Fig. \ref{Fig_extra}.}
This can be an artifact of the fitting procedure,
which can {yield} fitting ranges that include 
a piece of the bending-downwards part of the distribution
in the case of lemmas.
The only way to avoid this would be either to have infinite data, 
or not to find the fitting range automatically,
or to use a fitting distribution that parametrizes also the bending.
As we are {mostly} interested in the power-law regime, 
we have not considered these modifications to the fits.


A rescaling of the axes 
as in Refs. \cite{Peters_Deluca,Corral_csf}
can lead to additional support for {our} results
{(see also Ref. \cite{FontClos_Corral})}.
Fig. \ref{Fig2}(a) shows the rescaling for {\it La Regenta}.
Each axis is multiplied by a constant factor, in the form 
$$
\begin{array}{rll}
n &\rightarrow & n \langle n \rangle / \langle n^2 \rangle\\
f(n) &\rightarrow & f(n) \langle n^2 \rangle^2 / \langle n \rangle^3,\\
\end{array}
$$
which translates into a simple shift of the curves in a double-logarithmic plot,
not affecting the shape of the distribution and therefore 
keeping the possible power-law dependence.
The collapse of the tails of the two curves into a single one
is then {an alternative visual} indication of the stability of the exponents.
The results {for the 5 texts that were not shown before are now
displayed} in Fig. \ref{Fig2}(b).
These findings suggest that, in general, Zipf's law fulfills a kind of invariance
under lemmatization, 
{at least approximately},
although there can be exceptions {for some texts}.

Finally, in order to test the influence of the stream of consciousness part of {\it Ulysses}
on the results, we have repeated the fits removing that part of the text.
This yields a new text that is about 9\% shorter, but more homogeneous.
The Zipf exponents turn out to be $\gamma_w =1.98 \pm 0.01$ for $n\ge 6$
and $\gamma_l=2.02 \pm 0.04$ for $n\ge 32$, slightly higher than for the complete text.
Nevertheless, the new $\gamma_w$ and $\gamma_l$ still are compatible
between them (in the sense explained above for individual texts), 
and therefore our conclusions do not change regarding the
similarity between word and lemma exponents.
If we pay attention to the removed part, despite its peculiarity, 
the stream of consciousness prose still fulfills Zipf's law,
but with smaller exponents,
$\gamma_w=1.865 \pm 0.02$ for $n \ge 2$ and $\gamma_l=1.82 \pm 0.03$ for $n\ge 3$. 
Both exponents are also compatible between them.

\begin{figure}[!ht]
\begin{center}
\includegraphics[width=5in]{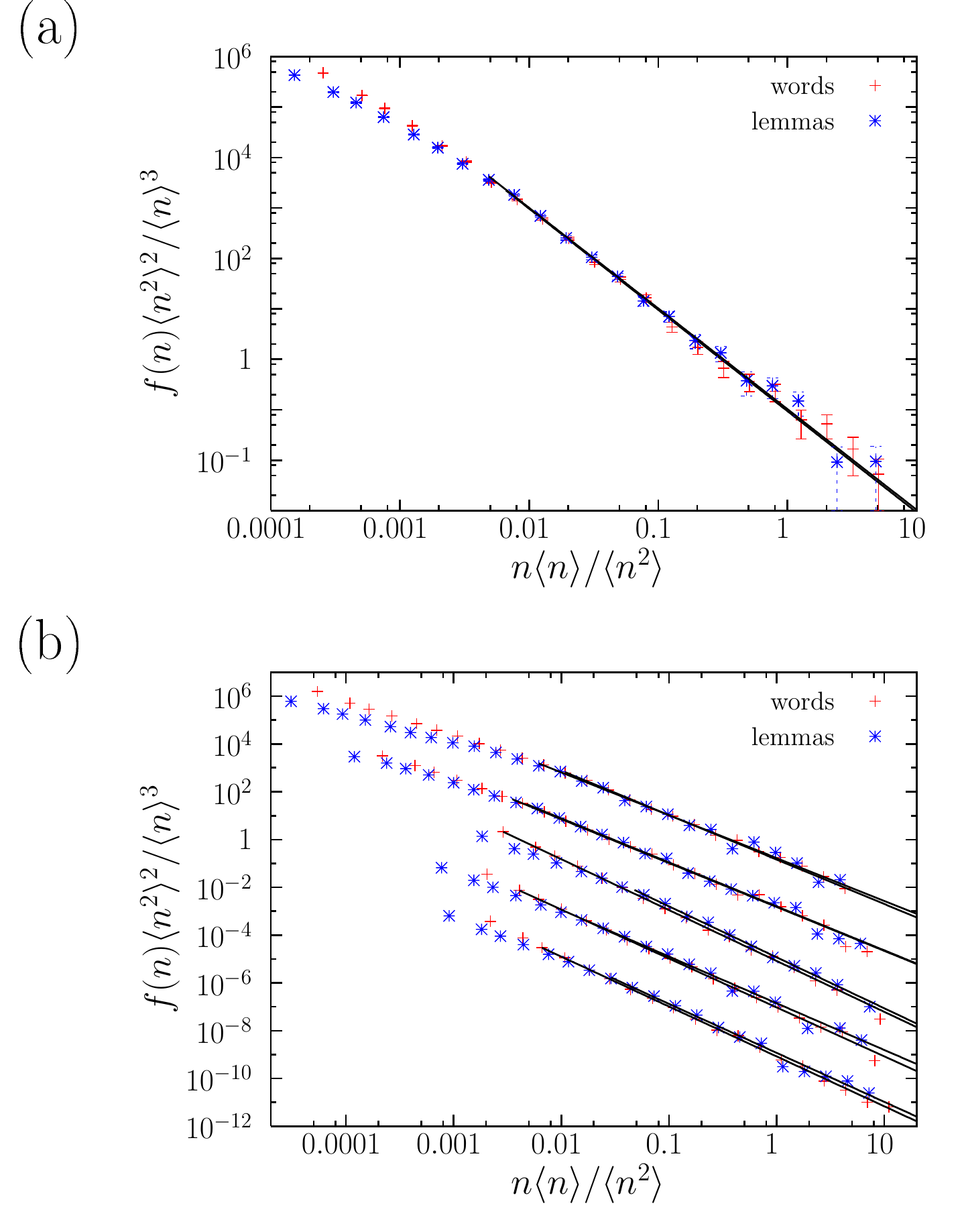}
\end{center}
\caption{
{(a) Probability mass functions $f(n)$ of the absolute frequencies $n$ of
words and lemmas in {\it La Regenta}, together with their fits,
under rescaling of both axis.
The collapse of the tails indicates the compatibility of both power-law
exponents.
(b) The same for, from top to bottom, 
{\it Artam\`ene, Bragelonne}
(both in French), 
{\it Seitsem\"an v., Kev\"at ja t.,} and {\it Vanhempieni r.}
(all three in Finnish).
The rescaled distributions are multiplied in addition by factors 
$1$, $10^{-2}$, etc., for a clearer visualization.
}  
}
\label{Fig2}
\end{figure}

\subsection*{{The consistency of the lower cut-offs of frequency for word forms and lemmas}}

As we {have done} with the exponent $\gamma$, 
we define $a_w$ and $a_l$ as the lower cut-off of {the power-law fit for the frequency distributions of
words and of lemmas, respectively. {Those values are compared in Fig. \ref{Fig_extrados}.}
{When all texts are considered,} 
{a Student $t-$test for paired samples
yields the rejection of the hypothesis that there
is no significant difference in the values of $a_w$ and $a_l$,
even if the presence of one possible outlier is taken into account
($t=-3.091$ and $p=0.015$).
In fact,}
$a_w$ and $a_l$ are not independent, as their Pearson correlation is $\rho  = 0.961$ 
($\mathcal{N}=10$ and {$p = 0.0014$  
for the null hypothesis $\rho =0$}, calculated through permutations of one of the variables).
{These results are not very surprising upon inspection of Fig. \ref{Fig_extrados}: Most points $(a_w, a_l)$ lay above the diagonal, suggesting a tendency for $a_l$ to exceed $a_w$. }

Like we did {for} the exponents, 
we formulate three hypotheses about the relationship between the low-frequency cut-offs. 
{The first hypothesis is that $a_w$ and $a_l$ are identically distributed for a given text.}
The second hypothesis is that the expectation of $a_w$ given $a_l$ is $E[a_w|a_l] = a_l$,
while the third hypothesis is {that} the expectation of $a_l$ given $a_w$ is $E[a_l|a_w] = a_w$. 
The second and third hypotheses are supported by the strong Pearson correlation between 
$a_w$ and $a_l$ {just mentioned}.
We define $\bar  a_w  $ and $\bar  a_l  $ as the mean value of $a_w$ and $a_l$, 
respectively, in our sample of  ten {texts}. 
The coin flipping test reveals that $|\bar  a_w   - \bar  a_l | = 16.9$ 
is significantly high 
($p$-value = $0.01$). 
Therefore, the first hypothesis does not stand, 
{not even after the {exclusion} of one outlier (which leads to $p=0.03$).}}

{The second hypothesis is indeed that $a_w/a_l$ is mean independent of $a_l$ 
while the third hypothesis is that $a_l/a_w$ is mean independent of $a_w$. 
Table \ref{mean_independence_table} indicates that neither a Pearson nor a Spearman correlation test are able to reject the second hypothesis. In contrast, a Pearson correlation test fails to reject the third hypothesis but the Spearman correlation test does reject it. {This should not be interpreted as an contradiction between Pearson and Spearman tests but as an indication that the relationship between $a_l$ and $a_w$ is non-linear, as suggested by Fig. \ref{Fig_extrados}}. As a typical correlation test is conservative because it only checks a necessary condition for mean dependence \cite{Ferrer2012h}, a further test is required.  
The second hypothesis states that 
$E[a_w |a_l] = c_1 a_l + c_2$ with $c_1 = 1$ and $c_2 = 0$ 
while the third hypothesis states that 
$E[a_l |a_w] = c_3 a_w + c_4$ with $c_3 = 1$ and $c_4 = 0$. 
A standard linear regression indicates that $c_1,c_3 \approx 1$ 
but 
$c_2\approx 0$ is in the limit of rejection,
whereas $c_4\approx 0$ fails
(Table \ref{linear_regression_table}). 
Therefore, this suggests that the cut-offs do not follow hypothesis 3.
{Note that the significance of the values $c_2 < 0$ and $c_4 > 0$ 
implies that, in general, 
$a_l$ is significantly larger than $a_w$.}} {This is consistent with Fig. \ref{Fig_extrados}.}

\begin{figure}[!ht]
\begin{center}
\includegraphics[width=5in]{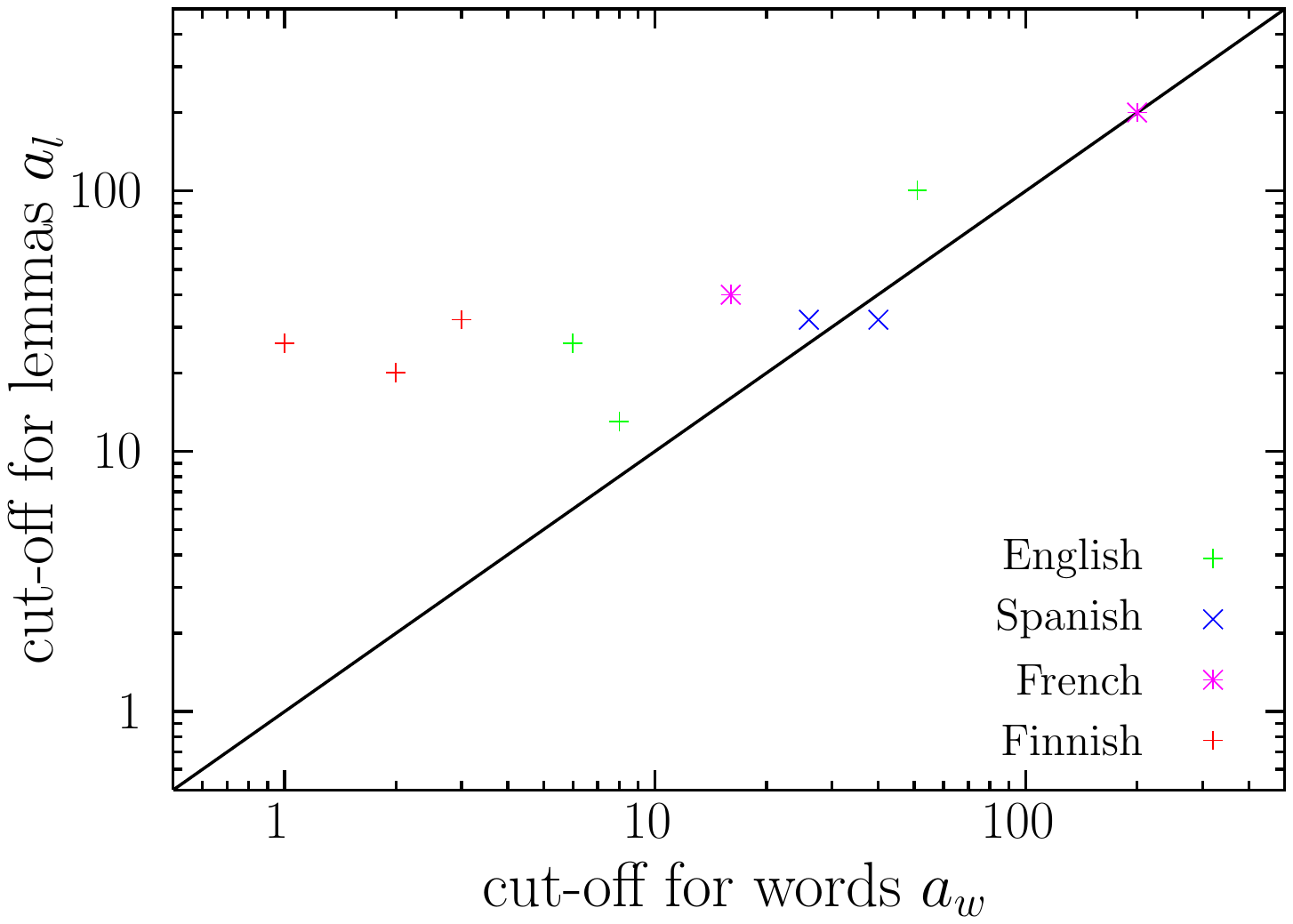}
\end{center}
\caption{
The lower cut-off for the frequency distribution of lemmas ($a_l$) versus the lower cut-off for the frequency distribution of word forms ($a_w$). The line $a_l = a_w$ is also shown (solid line). 
}
\label{Fig_extrados}
\end{figure}


%


\section*{Discussion}



{We have shown that Zipf's law is fulfilled {in long literary texts} for several orders of magnitude
in word and lemma frequency. The exponent of lemmas and the exponent of word forms are positively correlated. Similarly, the {low-frequency} cut-offs of lemmas and that of word forms are positively correlated. However, the exponent is more stable than the cut-off under the lemmatization transformation. 
While the exponent of lemmas is apparently centered around that of word forms and vice versa, the equivalent relationships are not supported for the cut-offs. 
{However, we cannot exclude the possibility that the exponents of lemmas are indeed not centered around those of word forms. Some suspicious evidence comes from Fig. \ref{Fig_extra}, where it can clearly be seen that $\gamma_l \leq \gamma_w$ in most cases. The tendency to satisfy this inequality is supported by the slight increase of the exponent $\alpha$ when moving from words to lemmas that has been reported in previous research \cite{Kwapien,Bentz2014} and that we have reviewed in the Introduction. Although Refs. \cite{Kwapien,Bentz2014} employed methods that differ substantially from ours, Eq. (\ref{exponents_equation}) allows one to interpret, with some approximation, the increase from $\alpha_w$ to $\alpha_l$ of Refs. \cite{Kwapien,Bentz2014} as the drop from $\gamma_w$ to $\gamma_l$ we have found in most cases.
} The apparent stability of the exponent of Zipf's law could be a type II error caused by the current size of our sample of long single-author texts. Furthermore, } the apparently constant relationship between $\gamma_l/\gamma_w$ and $\gamma_w$ (or between $\gamma_w/\gamma_l$ and $\gamma_l$) may hide a non-monotonic dependence, which the correlation tests above are blind to (our correlation tests are biased towards the detection of monotonic dependences).}}
{In spite of these limitations, one conclusion is clear: Exponents are more stable than cut-offs.} 
 
The similarity 
between the exponents of words and lemmas 
would be trivial if the lemmatization process affected only a few words,
or if these words were those with the smallest values of the frequency
(where the two distributions are more different).
However, Fig.~\ref{Fig_word_lemma}(a) 
displays the number of words that corresponds to each lemma
for {\it La Regenta} and for {\it Vanhempieni romaani} (in Finnish), 
showing that the effect of lemmatization is rather important \cite{Popescu_Altmann}.
Lemmatization affects all frequency scales, and, in some cases,
almost 50 words are assigned to the same lemma
in Spanish (verb paradigms), and more than 100 in Finnish (lemma {\it olla}).
All texts in Spanish, French, and Finnish yield very similar plots;
texts in English lead to flatter plots,
because lemmatization
is not such a big transformation there
due to the morphological characteristics of English. 
Fig.~\ref{Fig_word_lemma}(b) shows the same effect in a different way,
depicting the frequency of each word as a function of the frequency of its 
corresponding lemma. 
The presence of data above the diagonal is due to the fact that
some words can be associated to more than one lemma,
and then the sum of the frequencies of the words corresponding
to one lemma is not the frequency of the lemma;
{this is the case {in English} of the word
{\it found}, which can correspond to two lemmas, {\it (to) found}
or {\it (to) find}.}

\begin{figure}[!ht]
\begin{center}
\includegraphics[width=5in]{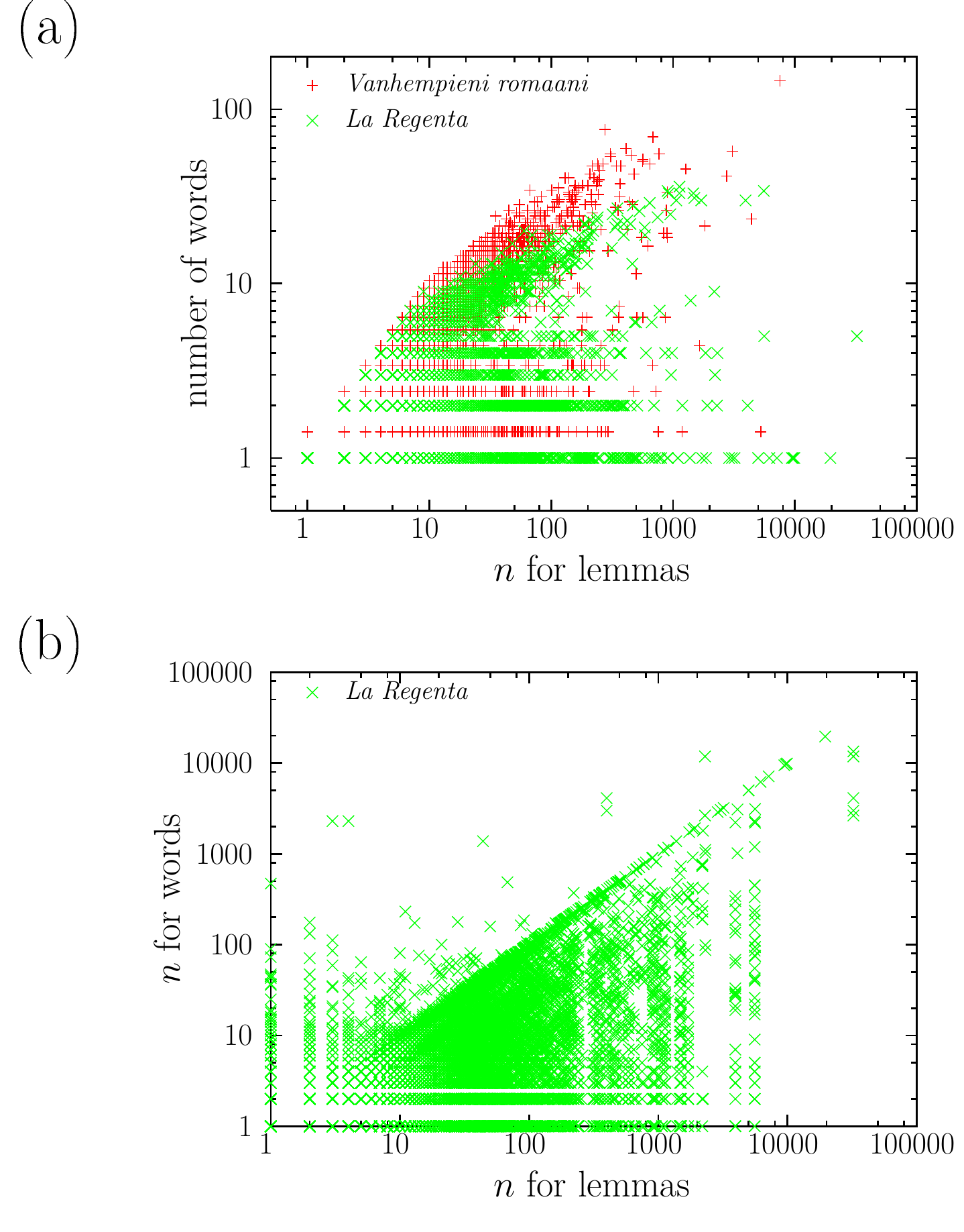}
\end{center}
\caption{
(a) Number of words per lemma as a function of lemma absolute frequency $n_l$
in {\it Vanhempieni romaani} (in Finnish) and in {\it La Regenta}.
The figures for the former have been slightly shifted up for clarity sake.
(b)
Frequency of words $n_w$ as a function of the frequency of their lemmas $n_l$
in {\it La Regenta}. 
}
\label{Fig_word_lemma}
\end{figure}


\begin{figure}[!ht]
\begin{center}
\includegraphics[width=5in]{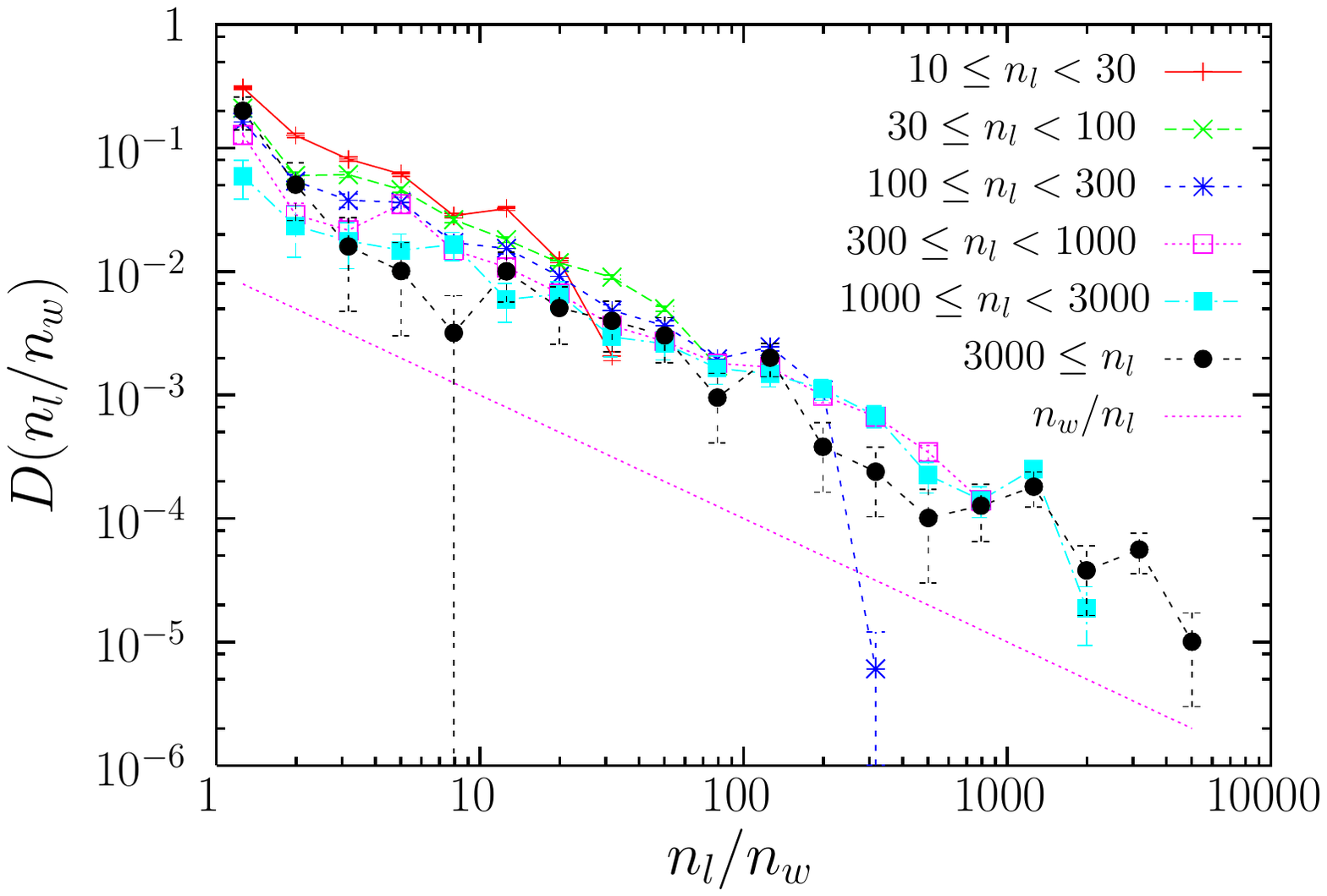} 
\end{center}
\caption{
Probability density $D(n_l/n_w)$
of the frequency ratio for lemmas and words, $n_l/n_w$,
in {\it La Regenta}.
Values of $n_l$ smaller than $n_w$ 
are disregarded, as they arise from words associated 
to more than one lemma.
Bending for the largest $n_l/n_w$ is expected as the maximum
of the ratio is given by $n_l$,
which is not constant for each distribution
but has a variation of half an order of magnitude 
(see {plot} legend).
}  
\label{Fig_Dratio}
\end{figure}

Finally, a complementary view is provided 
in Fig.~\ref{Fig_Dratio},
which shows the distribution of
the ratio of frequencies $n_l/n_w$ for the words
that correspond to a given lemma
(the subindices refer to lemmas and words, respectively).
In all cases this ratio is broadly distributed,
resembling a power law, although the statistics is too poor
to draw more solid conclusions.
As an indication, we plot in the figure a power law 
with exponent around 1, which is a good visual guide
for texts in Spanish and French.
In Finnish, the distribution becomes broader, 
being closer to a power law with exponent 0.5,
whereas in English the decay is faster, 
around an exponent 1.5 (not shown).
{In any case}, the relation between the frequency of words
and the frequency of their lemmas seems to lack
a characteristic scale.
The simplest case in which there is
only one word {per} lemma (and then their frequencies are the same,
$n_l/n_w=1$)
is quantified in {the last column of} Table \ref{Tabletwo}. 


{A challenge for future research is to illuminate the approximated invariance in the word-lemma transformation. 
A simplistic approach is offered by MacArthur's broken-stick model for species abundances \cite{MacArthur1957a}.}
Assume that each lemma, with frequency $n$,
only ``breaks'' into two different words, 
with frequencies in the text given by $m$ and $n-m$.
If $m$ is distributed uniformly between 0 and $n$, 
and the distribution of lemma frequencies is a power law, 
then, the distribution of word frequencies $m$ turns out 
to be also a power law, with the same exponent
(see the supplementary information of Ref. \cite{Corral_hurricanes}).
However, there is a long way from this oversimplification to reality. We have learned in Fig. \ref{Fig_word_lemma}(a)
that the number of words {a lemma} can yield varies a lot, from a few words for nominal lemmas to many words for verb lemmas in Spanish or French. 
More realistic models from an evolutionary perspective certainly appear as avenues for future work.



\section*{Conclusions}

{We have studied the robustness of Zipf's law under lemmatization 
for single-author written texts.
For this purpose it is crucial to unambiguously determine the power-law
exponent  $\gamma$
of the frequency distribution of types, and the range of validity
of Zipf's law, given by the low-frequency {cut-off} $a$, both for unlemmatized texts 
(consisting of word forms) and for lemmatized texts (transformed into sequences of lemmas).
We find that word and lemma distributions are somewhat different,
but the exponents of Zipf's law {in both cases remain close to each other},
for most of the texts,
{especially when compared to cut-offs}.  
{Nevertheless, the set of values of $\gamma$ 
suggests a slight bias for the exponents of lemmas to decrease with respect to that of words.}
%
{In contrast to the exponents, the cut-offs we find are not stable at all
  under the lemmatization transformation, 
  but are significantly increased, which in turn implies
 a decrease in the range of validity of Zipf's law}.
  Random breaking of lemmas into words {might} explain the relative stability of the power-law
distribution under the lemma-word transformation, but cannot account for the
wider validity of Zipf's law for words.

As Zipf's law is a paradigm that goes beyond linguistics,
having been found in the distribution of number of city inhabitants 
 \cite{Malevergne_Sornette_umpu}
or in the size of companies \cite{Axtell}
(among many other systems in which ``tokens'' merge to constitute ``types'' \cite{Clauset}),
our results could have a much broader applicability.
In many of these cases, 
the aggregation of tokens to form types can be done in different ways, or
types can be merged themselves to constitute ``supertypes'',
in a coarse-grained process akin both to lemmatization and to a transformation of the renormalization group
\cite{Corral_prl.2005}.
This is what was attempted in Refs. \cite{Jiang2011,Jiang2014}, 
where the spatial extent of elementary patches was added to define
what was called there a natural city.
Extrapolating our results, we could expect
that Zipf's exponent for city areas 
would not be very much affected by this process; 
in that case, the changes in Zipf's $\alpha$ exponent found in Ref. \cite{Jiang2011}
indicate that further study is necessary 
to elucidate whether the differences arise from the data (and so are due to differences in the underlying phenomenon) or from the data manipulation, 
e.g. 
the fitting method.
In general, investigating the commonalities and differences between different systems displaying Zipf's law is an area that should be actively addressed in the near future.


}

}

\section*{Materials and Methods}


\subsection*{Corpus selection}

First, we selected languages we have some command of (for data and error analysis purposes) 
and there are freely available lemmatization tools for \cite{Freeling,TreeTagger}. 
The exception is Finnish, which we included because it is a morphologically rich language that could shed light on the impact of lemmatization processes in Zipf's law.
We were interested in finding very long texts by single authors, 
and with that purpose we searched for the longest literary
texts ever written.
Of those novels published by mainstreaming publishers, 
{\it Artam\`ene} is ranked as the longest, 
in any language, and {\it Clarissa} as the longest in English
\cite{Wikipedia}.
{\it Don Quijote}, 
consistently considered the best literary piece ever written in Spanish,
is also of considerable length. 
The list was completed based on the  
availability of an electronic version of the novels in the 
{\it Project Gutenberg} \cite{Gutenberg}.
Note that {\it Artam\`ene} was not found in the {\it Gutenberg Project} but in a different source \cite{artamene}.
We were not able to find novels in Finnish of comparable length
to those in the other languages  
and in this case they are much shorter, see Table 1.

\subsection*{Lemmatization}

To carry out the comparison between word forms and lemmas, {texts must be lemmatized. 
A manual lemmatization} would have exceeded the possibilities of this project, 
so {we employed natural language processing} tools: 
{\it FreeLing} \cite{Freeling} for Spanish and English, 
{\it TreeTagger} \cite{TreeTagger} for French, 
and {\it Connexor}'s tools \cite{Connexor} for Finnish.

The tools carry out the following steps:

\begin{enumerate}

\item 	Tokenization: 
Segmentation of the texts into sentences and sentences into words, symbols, and 
punctuation marks (tokens).

\item	Morphological analysis: 
Assignment of one or more lemmas and morphological information ({a part-of-speech} tag) to each token. 
For instance, {\it houses} in English can correspond to the plural form of the noun 
{\it house} or the third person singular, present tense form of the verb {\it to house}. 
At this stage, both are assigned whenever the word form {\it houses} is encountered.

\item	Morphological disambiguation: An automatic tagger assigns the single 
most probable lemma and tag to each word form, depending on the context. 
For instance, in {\it The houses were expensive} the tagger would assign the nominal 
lemma and tag to {\it houses}, while in {\it She usually houses him}, the verb lemma and tag would be preferred. We note that as in both cases the lemma is the same,
both occurrences would count in the statistics of the {\it house} lemma.
\end{enumerate}


{As all these steps are automatic, some} errors are introduced at each step. 
However, the accuracy of the tools is quite high 
(e.g., around 95-97\% {at the token level} for morphological disambiguation), 
such that a quantitative analysis based on the results of the automatic process can be carried out.
{Also note that step 2 is based on a pre-existing dictionary 
(of words, not of lemmas, also called a lexicon): 
only the words that are in the dictionary 
are assigned a reliable set of morphological tags and lemmas. 
Although most tools {use heuristics to} assign tag and/or lemma 
information to words that are not in the dictionary, 
the results shown in this paper are obtained by counting
only tokens of lemmas for which the corresponding word types are found in the dictionary, 
so as to minimize the amount of error 
introduced by the automatic processing. 
This comes at the expense of losing some data. 
However, the dictionaries have quite a good coverage of the vocabulary, particularly at the token level, 
but also at the type level (see Table \ref{Tablethree}). 
The exceptions are {\it Ulysses}, because of the stream of consciousness prose, 
which uses many non-standard word forms, and {\it Artam\`ene}, 
because 17th century French contains many word forms 
that a dictionary of modern French does not include.



Note that the tools we have used do not only provide lemmatization, but also morphological analysis. 
That means that words are associated with a lemma ({\it houses}: {\it house}) and a morphological tag ({\it houses}: NNS, for {\it common noun in plural form}, or VBZ, for {\it verb in present tense, third person singular}). Tags express the main part of speech (POS; for {\it houses}, in this case, {\it noun} vs. {\it verb}) plus additional morphological information such as number, gender, tense, etc. That means that instead of reducing our vocabulary tokens to their lemmas, we could have 
chosen to reduce
them to their lemma plus tag information (lemma-tag, {\it house-NNS} vs. {\it house-VBZ}), or to their lemma plus POS information (lemma-POS: {\it house-N} vs. {\it house-V}). Table 6 shows that, from all these reductions, pure lemmatization ({\it houses}: {\it house}) is the most aggressive one, while still being linguistically motivated, as it reduces the size of vocabulary $V$ a factor which is between 2 (for {\it Moby-Dick}) 
and 5 (for {\it Artam\`ene}). 
Therefore, in this paper we focus on comparing word tokens with lemmas. 
%
%
%
%
A further reduction in the lemmatization transformation is provided by our requirement,
explained in the previous paragraph,
that the corresponding word is included in the dictionary of the lemmatization software.
If this restriction 
is eliminated, the results are very similar, 
as the restriction mainly operates at the smallest frequencies
(let us say, $n\le 10$ or $20$), whereas the power law fit takes place
for larger frequencies (see Table \ref{Tabletwo}).
Alternatively to lemmatization, there is a different transformation that, instead of aggregating
words into lemma-POS or lemmas, segregates words into what we may call word-lemma-tag.
Table \ref{Tablelast} shows that this transformation is not very significant, 
in terms of changes in the size of the vocabulary.

\subsection*{Statistical procedures}

{We now explain the different statistical tools used in the paper.
We begin with the
procedure to find {parameter values}
that describe the distributions of frequencies, 
that is, the power-law exponent $\gamma$ and the low-frequency cut-off $a$.
{As we have already mentioned, the method we adopt is based on the one by
Clauset {et al.} \cite{Clauset},
but it incorporates important modifications that have been shown to yield
a better performance in the continuous case \cite{Peters_Deluca,Corral_nuclear}.}
The algorithm we use is the one described in Ref. \cite{Corral_Deluca_arxiv}.

The key issue when fitting power laws is to {determine} the optimum value $a$ of the variable
for which the power-law fit holds. The method starts by selecting arbitrary values of $a$,
and for each value of $a$ the maximum likelihood estimation of the exponent is
obtained. In the discrete case one has to {maximize} the likelihood
function numerically, where the normalization factor is obtained from the Hurwitz zeta function.
The goodness of the fit needs to be evaluated independently.
For this, the method uses the Kolmogorov-Smirnov test, and the $p$-value of the fit
is obtained from Monte Carlo simulations of the fitted distribution.
The simulated data need to undergo the same procedure as
the original empirical data in order to avoid biases in the fit
(which would lead to inflated $p$-values).
In this way, for each value of $a$ we obtain a fit 
and a quantification of the goodness of the fit given by its $p$-value.
The chosen value of $a$ is the smallest one (which gives the largest
power-law range), provided that its $p$-value is large enough. 
{This has an associated estimated maximum likelihood exponent, which is the
final result for exponent. Its standard deviation (for the quantification of its uncertainty) is obtained, 
for fixed $a$, from the standard deviation of the values obtained in the Monte Carlo simulations.}

The complete algorithm is implemented here with the following specifications.
The minimum frequency $a$ is sampled with a resolution of 10 points per 
{order of magnitude,} 
{in geometric progression to yield a constant
separation of $a-$values in logarithmic scale. {The procedure is simple: 
A given value for $a$ is obtained by multiplying its previous value by 
$\sqrt[10]{10} \approx 1.26$, with the initial value of $a$ being $1$,}
and in this sense the relative error in $a$ can be considered to be
of the order of $100(\sqrt[10]{10}-1) \approx 26 \%$; the values of $a$ produced that are not integers are 
{rounded to the next integer} {\em a posteriori} to become true parameters.}
The goodness of fit is evaluated with 1000 Monte Carlo
simulations; and a {$p$-value is considered to be large enough if it exceeds $0.20$}.

{Now we review the methods used to investigate {the similarity between words and lemmas from the perspective of the parameters of the frequency distribution.} 
Student's $t-$test for paired samples
makes use of the differences between the values of the parameters of each text
(either exponents or cut-offs, word minus lemma) and rescales the mean of the differences by dividing it by the
(unbiased) standard deviation of the differences and by multiplying by $\sqrt{\mathcal{N}}$ (with $\mathcal{N}$ the number of data,
10 {books} in our case). 
This yields the $t$ statistic, which, if the differences are normally distributed
with the same standard deviation and zero mean, follows a $t-$Student distribution with 
$\mathcal{N}-1$ degrees of freedom.
Simulations of $\mathcal{N}$ independent normally distributed variables
with zero mean and the same standard deviation
mimic the distribution of the differences under the null hypothesis
and lead to the $t-$Student distribution,
which allows the calculation of the $p-$value.
This simulation method allows for the systematic treatment of outliers, 
as mentioned in the main text ({if one outlier is removed}, then, obviously, $\mathcal{N}=9$
in the calculation of {the} value of $t$).

Correlations between parameters 
are calculated using either the Pearson 
correlation coefficient or the Spearman correlation coefficient. {While the Pearson coefficient is a measure of the strength of the linear association, the Spearman correlation coefficient is able to detect non-linear dependences \cite{Conover1999a,Zou2003a}}. 
The former is defined as the covariance divided by the product of
the standard deviations; the latter is defined in the same way
but replacing the values of each variable by their ranks (one, two, etc.);
both are represented by $\rho$.
In order to test the null hypothesis $\rho =0$ we perform a reshuffling of one of
the variables and calculate the resulting $\rho $. The $p$-value is just
the fraction of values of $\rho $ for the reshuffled data with absolute value larger or equal
than the absolute value of $\rho $ for the original data
(a two-sided test). 
}

{We could have also used a correlation ratio test \cite{Ferrer2012h}, a test based on the correlation ratio, another correlation statistic \cite{Kruskal1958a}. That test provides a way of testing for mean independence that is {\em a priori} more powerful than a standard correlation test (a Pearson correlation test is a conservative test of mean dependence \cite{Ferrer2012h}). However, our dataset exhibits a high diversity of values (Table \ref{Tabletwo}), which is known to lead to type II errors with that statistic \cite{Ferrer2012h}.   
}

\section*{Acknowledgments}

We thank 
the \emph{Connexor Oy} company for tagging the Finnish texts, 
R. D. Malmgren and P. Puig for discussions about statistical testing, 
{G. Altmann for his advice from a quantitative linguistics perspective,}
L. Devroye for facilitating a copy of his valuable book for free,
and A. Deluca and F. Font-Clos for 
feedback on the fitting of word and lemma frequency distributions.

%





\section*{Tables}

\begin{table}[!ht]
\begin{center}
\caption{
Characteristics of the books analyzed.
The length of each book $L$ is measured 
in millions of tokens.
\label{Tableone}
}
\smallskip
\begin{tabular}{lllrc}
 Title & Author & Language & Year & $L$ \\
\hline
Clarissa$^1$ & Samuel Richardson & English & 1748 & $0.976  $  \\
Moby-Dick$^2$ & Herman Melville   & English & 1851  & $ 0.215  $  \\
Ulysses    & James Joyce   & English & 1918 & $  0.269  $  \\
Don Quijote$^3$ & Miguel de Cervantes & Spanish & 1605   & $  0.381  $  \\
La Regenta & L. Alas ``Clar\'{\i}n'' & Spanish & 1884 & $  0.308  $  \\
Artam\`ene$^4$ & Scud\'ery siblings$^9$ & French & 1649 & $  2.088  $  \\
Le Vicomte de Bragelonne$^5$ & A. Dumas (father)  & French & 1847  & $ 0.699  $  \\
Seitsem\"{a}n veljest\"{a}$^6$ & Aleksis Kivi & Finnish & 1870 & 0.081 \\
Kev\"{a}t ja takatalvi$^7$ & Juhani Aho & Finnish & 1906& $0.114  $ \\
Vanhempieni romaani$^8$  & Arvid J\"{a}rnefelt & Finnish & 1928&  0.136 \\
\end{tabular}
\end{center}
$^1$Clarissa: Or the History of a Young Lady. 
$^2$Moby-Dick; or, The Whale.
$^3$El ingenioso hidalgo don Quijote de la Mancha (1605) --
The Ingenious Gentleman Don Quixote of La Mancha (title in English);
including second part:
El ingenioso caballero don Quijote de la Mancha (1615).
$^4$Artam\`ene ou le Grand Cyrus -- Artam\`ene, or Cyrus the Great.
$^5$Le Vicomte de Bragelonne ou Dix ans plus tard --
The Vicomte of Bragelonne: Ten Years Later.
$^6$Seven Brothers.
$^7$Spring and the Untimely Return of Winter.   
$^8$The Story of my Parents.
$^9$Madeleine and Georges de Scud\'ery.
\end{table}

\begin{table}[!ht]
\caption{
Power-law fitting results for words and lemmas, denoted respectively
by {subindices} $w$ and $l$. 
{$V$ is the number of types (vocabulary size),
$n_m$} {is}
{the maximum frequency of the distribution,
$N_a$ {is} the number of types in the power-law tail, 
i.e., with $n \ge a$,
$a$ is the minimum value for which the power-law fit holds,
and $\gamma$ and $\sigma$ are the power-law exponent and
its standard deviation, respectively.
} 
{$2\sigma_d$, the double of the standard deviation $\sigma_d$ is also given.
$\sigma_d$ is the standard deviation {of} $\gamma_l - \gamma_w$ assuming independence, which is $\sigma_d=\sqrt{\sigma_w^2+\sigma_l^2}$}.
The last column provides $\ell_1$, the number of lemmas associated to only one word form.
Notice that the lemma exponent is very close to the one found in Ref. \cite{FontClos_Corral}
for the tail of a double power-law fitting, except for {\it Moby-Dick} and {\it Ulysses}.
}
\smallskip
\hspace{-1cm}
\begin{tabular}{lrrrrrrrrrrrr}
Title & $V_w$ & $n_{mw}$ & $N_{a_w}$ & $a_w$ & $\gamma_w \pm \sigma_w$ 
      & $V_l$ & $n_{ml}$ & $N_{a_l}$ & $a_l$ & $\gamma_l \pm \sigma_l$ & $2 \sigma_d$ & $\ell_1$\\
\hline
%
Clarissa      &   20492 &38632 & 1514 &    51 &  1.83$\pm$0.02    &
                   9041 &41679 & 838 &   101 &  1.83$\pm$0.03    &  0.07 & 5750  \\
Moby-Dick     &   18516 &14438 & 2658 &     8 &  1.97$\pm$0.02    &
                   9141 &14438 & 1548 &   13 &  1.90$\pm$0.02    &  0.06 & 6157\\
Ulysses       &   29450 &14934 & 4377 &     6 &  1.95$\pm$0.01    &
                  12469 &14934 & 1024 &  26 &  1.97$\pm$0.03    &  0.07 & 8670\\
Don Quijote   &   21180 &20704 & 939 &   40 &  1.93$\pm$0.03    &
                   7432 &31521 & 936 &  32 &  1.83$\pm$0.03    & 0.08 &  3812 \\
La Regenta    &   21871 &19596 & 1196&   26 &  2.01$\pm$0.03    &
                   9900 &32300 & 993 &  32 &  2.00$\pm$0.03    &  0.08 & 5308 \\
Artam\`ene    &   25161 & 88490& 936 &   200 &  1.86$\pm$0.03    &
                   5008 &119016& 641 & 200 &  1.79$\pm$0.03    & 0.08 &  2178 \\
Bragelonne    &   25775 & 26848& 3173 &   16 &  1.84$\pm$0.02    &
                  10744 & 45577& 1382 &  40 &  1.84$\pm$0.02    &  0.06 & 5391 \\
Seitsem\"an      &22035 &  4247& 22035 &   1 &  2.13$\pm$0.01    &
                   7658 &  4247& 474 &  26 &  2.13$\pm$0.05    &  0.10 & 4246 \\
Kev\"at ja    &   25071 &  5042& 8660 &    2 &  2.05$\pm$0.01    &
                   8898 &  6886& 699 &  20 &  1.96$\pm$0.04    &  0.07 & 5060 \\
Vanhempieni   &   35931 &  5254& 6523 &     3 &  2.09$\pm$0.01    &
                  13510 &  7526& 571 &  32 &  2.05$\pm$0.04    &  0.09 & 7837 \\
\end{tabular}
\begin{flushleft}
\end{flushleft}
\label{Tabletwo}
\end{table}

\begin{table}[!ht]
\begin{center}
\caption{\label{mean_independence_table} 
Analysis of the association between random variables 
using Pearson and Spearman correlations as statistics. 
$\rho $ is the value of the correlation statistic and 
$p$ is the $p$-value of a two-sided test
with null hypothesis $\rho =0$, calculated through permutations of one 
of the variables (the results can be different if $p$ is calculated from a $t-$test).
The sample size is $\mathcal{N}=10$ in all cases.
{Only the Spearman correlation between $a_w$ and $a_l/a_w$ is significantly
different from zero.}
}
\smallskip
\begin{tabular}{llll}
Association & Correlation test & $\rho $ & $p$ \\ \hline
$\gamma_w/\gamma_l$ and $\gamma_l$ & Pearson correlation test  & $-0.378$  & $0.28$ \\
                                & Spearman correlation test & $-0.418$  & $0.23$ \\
$\gamma_l/\gamma_w$ and $\gamma_w$ & Pearson correlation test  & $-0.034$ & $0.92$ \\
                                & Spearman correlation test & $-0.091$ & $0.81$ \\
$a_w/a_l$ and $a_l$ & Pearson correlation test & $\phantom{-}0.420$ & $0.24$ \\
                    & Spearman correlation test & $\phantom{-}0.393$ & $0.26$ \\  
$a_l/a_w$ and $a_w$ & Pearson correlation test  & $-0.373$ & {$0.11$} \\ 
                    & Spearman correlation test & $-0.867$ & $0.002$ \\
\end{tabular}
\end{center}
\end{table}

\begin{table}[!ht]
\begin{center}
\caption{\label{linear_regression_table} 
The fit of a linear model for the relationship between exponents 
($\gamma_w$ and $\gamma_l$) and the relationship between cut-offs ($a_w$ and $a_l$). 
$c_1$ and $c_3$ stand for slopes and $c_2$ and $c_4$ stand for intercepts. 
The error bars correspond to one standard deviation.
A {Student's} $t$-test is applied to {investigate} if the 
{slopes are} significantly different from  
{one and if the intercepts are significantly different from zero.
The resulting $p$-values indicate that
in all cases the slopes are compatible with
being equal to one. The intercepts are compatible with zero for the exponents, but
seem to be incompatible for the cut-offs.
}}
\smallskip
\begin{tabular}{llrrr}
Linear model                               & Parameters & Student's $t$ & $p$ \\ \hline
{$E[\gamma_w|\gamma_l] = c_1\gamma_l + c_2$}  & $c_1 = \phantom{-0}0.855 \pm 0.135$    & $-1.074$ & $0.314$ \\ 
                                           & $c_2 = \phantom{-0}0.315 \pm 0.261$    & $1.208$  & $0.261$ \\
$E[\gamma_l|\gamma_w] = c_3\gamma_w + c_4$ & $c_3 = \phantom{-0}0.975 \pm 0.154$    & $-0.161$ & $0.876$ \\ 
                                           & $c_4 = \phantom{-0}0.013 \pm 0.303$    & $0.044$  & $0.966$ \\
{$E[a_w|a_l] = c_1 a_l + c_2$}               & $c_1 = \phantom{-0}1.012 \pm 0.103$    & $0.115$ & $0.911$ \\ 
                                           & $c_2 = -17.523 \pm 7.798$  & $-2.247$ & $0.055$ \\
$E[a_l|a_w] = c_3 a_w + c_4$               & $c_3 = \phantom{-0}0.912 \pm 0.093$    & $-0.945$ & $0.372$ \\ 
                                           & $c_4 = \phantom{-}20.009 \pm 6.272$   & $3.190$  & $0.013$ \\
\end{tabular}
\end{center}
\end{table}

\begin{table}[!ht]
\begin{center}
\caption{
Coverage of the vocabulary by the dictionary in each language, 
both at the word-type and at the token level.
The average for all texts is also included.
{Remember that we
distinguish between a word {\it type} (corresponding to its orthographic form) 
and its {\it tokens} (actual occurrences in text). 
}
}
\smallskip
\begin{tabular}{lll}
 Title         & Tokens     & Types   \\
\hline
Clarissa       & 96.9 \% & 68.0 \%   \\
Moby-Dick    & 94.7 \%  & 70.8 \%   \\ 
Ulysses        & 90.4 \% & 58.6 \%    \\
Don Quijote    & 97.0 \% & 81.3 \%   \\
La Regenta      & 97.9 \%  & 89.5 \%  \\
Artam\`ene     & 83.6 \% & 43.6 \%    \\
Bragelonne      & 97.5 \% & 89.8 \%   \\
Seitsem\"an v.  & 95.4 \%  & 89.8 \%  \\
Kev\"at ja t.  & 98.3 \% & 96.2 \%    \\
Vanhempieni r.  & 98.5 \% & 96.5 \%   \\
average        & 95.0 \% & 78.4 \%    \\
\end{tabular}
\label{Tablethree}
\end{center}
\end{table}

\begin{table}[!ht]
\begin{center}
\caption{
Size of vocabulary $V$ (i.e., number of types) when texts are decomposed in different
sorts of types, being these: word-lemma-tag (w-l-t), plain words, lemma-POS (l-pos),
lemma-POS of words in the dictionary (l-pos dic), 
lemmas, and lemmas of words in the dictionary (lemma dic).
The latter provide the most radical transformation, 
as it yields the largest reduction in resulting vocabulary.
}
\smallskip
\begin{tabular}{lrrrrrr}
            & w-l-t &     word  & l-pos & l-pos dic & lemma & lemma dic \\ \hline
     Clarissa   &    23624  &    20492  &    17058  &    10315  &    15356  &     9041  \\
      Moby-Dick  &    20777  &    18516  &    15774  &    10426  &    14226  &     9141  \\
      Ulysses    &    32952  &    29450  &    26412  &    14136  &    24089  &    12469  \\
      Don Quijote&    23359  &    21180  &    11872  &     7906  &    11128  &     7432  \\
      La Regenta &    24053  &    21871  &    12509  &    10500  &    11768  &     9900  \\
      Artam\`ene &    31574  &    25161  &     7605  &     5349  &     7177  &     5008  \\
      Bragelonne &    28803  &    25775  &    12994  &    11342  &    12127  &    10744  \\
     Seitsem\"an&    22851  &    22035  &     9749  &     7788  &     9607  &     7658  \\ 
     Kev\"at ja &    26087  &    25071  &     9897  &     9054  &     9733  &     8898  \\
      Vanhempieni &    37247  &    35931  &    14751  &    13678  &    14566  &    13510  \\
\end{tabular}
\label{Tablelast}
\end{center}
\end{table}

\end{document}